\documentclass[aps,prb,twocolumn,shortbibliography,superscriptaddress]{revtex4-1}
\usepackage{epsfig}
\usepackage{epstopdf}
\usepackage{amsmath}
\usepackage{amsfonts}
\usepackage{amssymb}
\usepackage{hyperref}
\usepackage{bm}
\usepackage{makecell}
\usepackage{rotating}
\usepackage{hyperref}
\usepackage[utf8]{inputenc}
\usepackage[normalem]{ulem}

\usepackage{graphicx}% Include figure files
\usepackage{dcolumn}% Align table columns on decimal point
\usepackage{bm}% bold math
\usepackage{color}

\usepackage{tikz,xcolor,hyperref}

\definecolor{lime}{HTML}{A6CE39}
\DeclareRobustCommand{\orcidicon}{%
	\begin{tikzpicture}
	\draw[lime, fill=lime] (0,0)
	circle [radius=0.16]
	node[white] {{\fontfamily{qag}\selectfont \tiny ID}};
	\draw[white, fill=white] (-0.0625,0.095)
	circle [radius=0.007];
	\end{tikzpicture}
	\hspace{-2mm}
}

\foreach \x in {A, ..., Z}{%
	\expandafter\xdef\csname orcid\x\endcsname{\noexpand\href{https://orcid.org/\csname orcidauthor\x\endcsname}{\noexpand\orcidicon}}
}

\begin{document}

%\title{Coexistence of two altermagnetic spin-splittings in nonsymmorphic Pnma space group}  
% Title for Amar contract, to send to DEADLINE: Wojto before 20 June.

\title{Interplay between altermagnetism and nonsymmorphic symmetries generating \\ large anomalous Hall conductivity by semi-Dirac points induced anticrossings}

\author{Amar Fakhredine\orcidB}
\affiliation{Institute of Physics, Polish Academy of Sciences, Aleja Lotnik\'ow 32/46, PL-02668 Warsaw, Poland}
\affiliation{International Research Centre Magtop, Institute of Physics, Polish Academy of Sciences,
Aleja Lotnik\'ow 32/46, PL-02668 Warsaw, Poland}

\author{Raghottam M. Sattigeri\orcidC}
\affiliation{International Research Centre Magtop, Institute of Physics, Polish Academy of Sciences,
Aleja Lotnik\'ow 32/46, PL-02668 Warsaw, Poland}

\author{Giuseppe Cuono\orcidD}
\email{gcuono@magtop.ifpan.edu.pl}
\affiliation{International Research Centre Magtop, Institute of Physics, Polish Academy of Sciences,
Aleja Lotnik\'ow 32/46, PL-02668 Warsaw, Poland}

\author{Carmine Autieri\orcidA}
\affiliation{International Research Centre Magtop, Institute of Physics, Polish Academy of Sciences,
Aleja Lotnik\'ow 32/46, PL-02668 Warsaw, Poland}

\date{\today}
\begin{abstract}
We investigate the interplay between altermagnetic spin-splitting and nonsymmorphic symmetries using the space group no. 62 as a testbed.
Studying different magnetic orders by means of \textit{first-principles} calculations, we find that the altermagnetism (AM) is present in the C-type magnetic configuration while it is absent for the G-type and A-type configurations due to different magnetic space group types. The nonsymmorphic symmetries constrain the system to a four-fold degeneracy at the border of the Brillouin zone with semi-Dirac dispersion.
In the case of large hybridization as for transition metal pnictides, the interplay between AM and nonsymmorphic symmetries generates an intricate network of several crossings and anticrossings that we describe in terms of semi-Dirac points and glide symmetries.
When we add the spin-orbit coupling (SOC), we find a N\'eel-vector dependent spin-orbit splitting at the time-reversal invariant momenta points since the magnetic space groups depend on the N\'eel vector. The magnetic space group type-I produces antiferromagnetic hourglass electrons that disappear in the type-III. 
When the N\'eel vector is along x, we observe a glide-protected crossing that could generate a nodal-line in the altermagnetic phase.
The SOC splits the remaining band crossings and band anticrossings producing a large anomalous Hall effect in all directions excluding the N\'eel-vector direction.
\end{abstract}


\maketitle

\section{Introduction}

% altermagnetism

Until a few years ago, the two collinear magnetic phases were known in condensed matter physics as ferromagnetism and antiferromagnetism displaying completely different properties.
Very recently, a new variant of the collinear antiferromagnetism was discovered called altermagnetism (AM) or collinear antiferromagnets with non-interconvertible spin-structure motif pair\cite{Smejkal22,Smejkal22beyond,yuan2023degeneracy,hayami2019momentum,hayami2020bottom,Sanyal2023} which hosts both properties of the ferromagnets and usual antiferromagnets.
Unlike ferromagnetism, where the crystal has a net magnetization and there is time-reversal symmetry (TRS) breaking, and unlike antiferromagnetism, where the total magnetization is zero, altermagnetism hosts systems in which the magnetization in the real space is zero but there is breaking of the spin degeneracy in the reciprocal space like in ferromagnetic compounds. 
The condition to observe the AM is the absence of a translation, an inversion or a combination of both that maps the spin-up charge to the spin-down charge. In this case, only a roto-translation or mirror can map the spin-up charge in the spin-down charge. From the point of view of group theory, the altermagnetic compounds must belong to type-I and type-III magnetic space groups (MSG).\cite{GUO2023100991} The type-I MSGs are crystallographic space groups without any additional symmetry while the type-III are crystallographic space groups with additional antisymmetry versions of half of the symmetry operations.\cite{Bradley2009-bv}\\

The altermagnetic systems exhibit nonrelativistic spin-splitting and may produce anomalous Hall effect (AHE)\cite{doi:10.1126/sciadv.aaz8809,Feng2022} once the relativistic effects are included.
The AHE is enhanced by avoided crossings, also called anticrossings. While further investigations should be made in order to assure the possession of AHE , AM can produce an AHE along the direction of the Hall vector, but a very limited number of altermagnetic systems are metallic. Regarding technological applications,
altermagnets could be assumed as a leading role in realizing concepts in spincaloritronics\cite{zhou2023crystal}. They can be also used in Josephson junctions\cite{Ouassou23}, room-temperature magnetoresistance in an all-antiferromagnetic tunnel junction\cite{Qin2023} and to generate neutral currents for spintronics\cite{Shao21}.\\

% nonsymmorphic symmetries 
One of the space groups in which the altermagnetic phase was established is the Pnma space group\cite{GUO2023100991}. 
In a recent work, a new route to 
search altermagnetic states was introduced and it was taken in consideration the example of the Pnma perovskites \cite{Fernandes23}; it was shown that distinctive signatures on the band structure emerge from the angular variation of magnetization components in altermagnets. These signatures manifest as protected nodal lines along mirror planes of the crystal structure and pinch points on the Fermi surface, which act akin to type-II Weyl nodes \cite{Fernandes23}.
The Pnma presents several nonsymmorphic symmetries\cite{Zhao16,Watanabe16,Konig97}, which are a composition of fractional lattice translations with point-group operations, like mirror reflection (glide plane) or rotation (screw axis). The glide symmetry or glide reflection symmetry is a symmetry operation that consists of a combination of a mirror reflection with respect to a plane and then a translation parallel to that plane. The eigenvalues of the glide symmetry are +1 and -1, and the eigenvectors are orthogonal, therefore, the bands produce a protected band crossing.
In the absence of magnetism and SOC, the nonsymmorphic symmetries force the electronic bands to be degenerate at the borders of the Brillouin zone with the presence of semi-Dirac points\cite{Wieder16,Zhao16}, generating linear magnetoresistance\cite{Campbell2021}, unconventional topological phases\cite{Wang16Nature,Bzdusek16,Parameswaran13,Wieder16PRB,Brzezicki17, Yoshida19,Daido19} with unique surface states, Fermi surfaces with reduced dimensionality\cite{Cuono19PRM} and topological nonsymmorphic crystalline superconductivity with $\mathbb{Z}_4$ topological invariant\cite{Yoshida19,Daido19}. In the presence of SOC and magnetism, the system presents a partial or selective removal of the degeneracy\cite{Cuono19PRM}. The semi-Dirac points exhibit linear band dispersion in one direction (the high-symmetry direction at least in the space-group 62) and quadratic band dispersion in the orthogonal direction\cite{PhysRevB.93.125113}. The existence of these semi-Dirac points in the space group 62 and its relation with the nonsymmorphic symmetries has been widely demonstrated in literature \cite{Niu17,Niu19,Cuono19PRM,Campbell2021,Cuono19EPJST}. The connection between semi-Dirac points and glide symmetry will be described later in the text. \\

% CrAs
An example of a material with the Pnma crystal phase is the CrAs compound\cite{Wu10,Wu14} in the MnP-type phase, which is a rare itinerant antiferromagnet\cite{Wu10,Wu14} that could exhibit AHE due to AM.
%CrAs was also proposed to be a topological antiferromagnet\cite{Xu2020}, and it presents a superconducting phase under pressure in the vicinity of the magnetic phase, therefore it is an interesting material showing the interplay between topological, magnetic and superconducting properties as well as a probable unconventional type of superconductivity driven by the magnetic fluctuations\cite{Wu14}.
Additionally, the CrAs system is not ionic, therefore, $p$- and $d$-bands are strongly hybridizing with a large bandwidth of the order of 11 eV\cite{Autieri17CrAsPM}. This grants us the possibility to observe induced AM in the $p$-bands.
The magnetic ground state of the CrAs is a helimagnetic phase, with the components of the magnetic moment\cite{Wu10,Wu14,Shen16,Keller15} in the $a$-$b$ plane and with a spin-orbit coupling mostly from the $p$-states of As\cite{Autieri17CrAsPM,Cuono19EPJST,Wadge22}.
Here we investigate the CrAs as a testbed exclusively in hypothetical collinear magnetic orders since we are mainly interested in the interplay between nonsymmorphic symmetries and AM. Further investigations are necessary to extend these results to the non-collinear phase and to understand how much AM will survive in the non-collinear phase\cite{Zhu23}. CrAs will work as a prototype for other itinerant antiferromagnets of the same space group as MnPd$_2$, FeP, MnP or CuMnAs that may have a smaller AHE due to the lower spin-orbit coupling\cite{Campbell2021}. 
Different from most studied altermagnetic Pnma compounds until now\cite{Sattigeri23altermagnetic,Cuono23orbital,PhysRevB.107.155126} such as LaMnO$_3$, YVO$_3$ and CaCrO$_3$ that have their magnetic atoms in the Wyckoff positions 4b\cite{AUTIERI2023414407}, the atoms of CrAs are only present in the 4c Wyckoff positions, therefore, the system belongs to a different magnetic space group that should be more symmetric and host unexplored properties.

% short summary of our results. 

\begin{figure}[t!]
\centering
\includegraphics[width=4.19cm,angle=0]{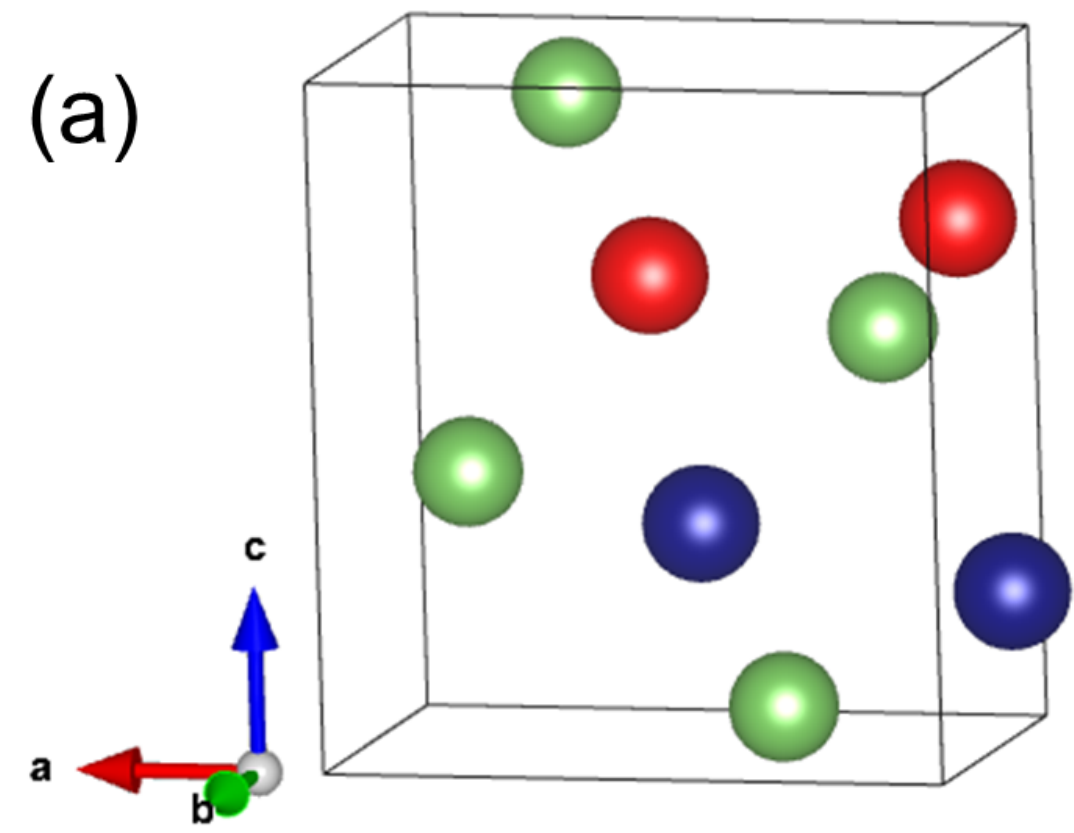}
\includegraphics[width=4.19cm,angle=0]{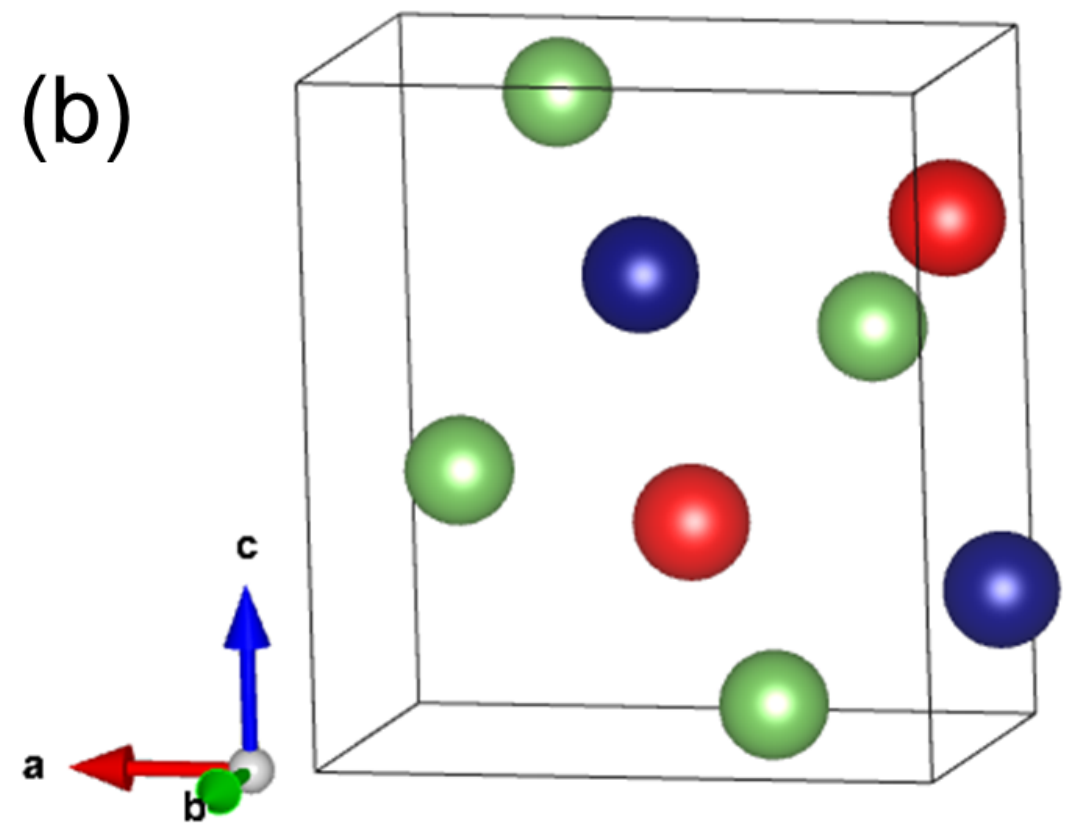}
\includegraphics[width=4.19cm,angle=0]{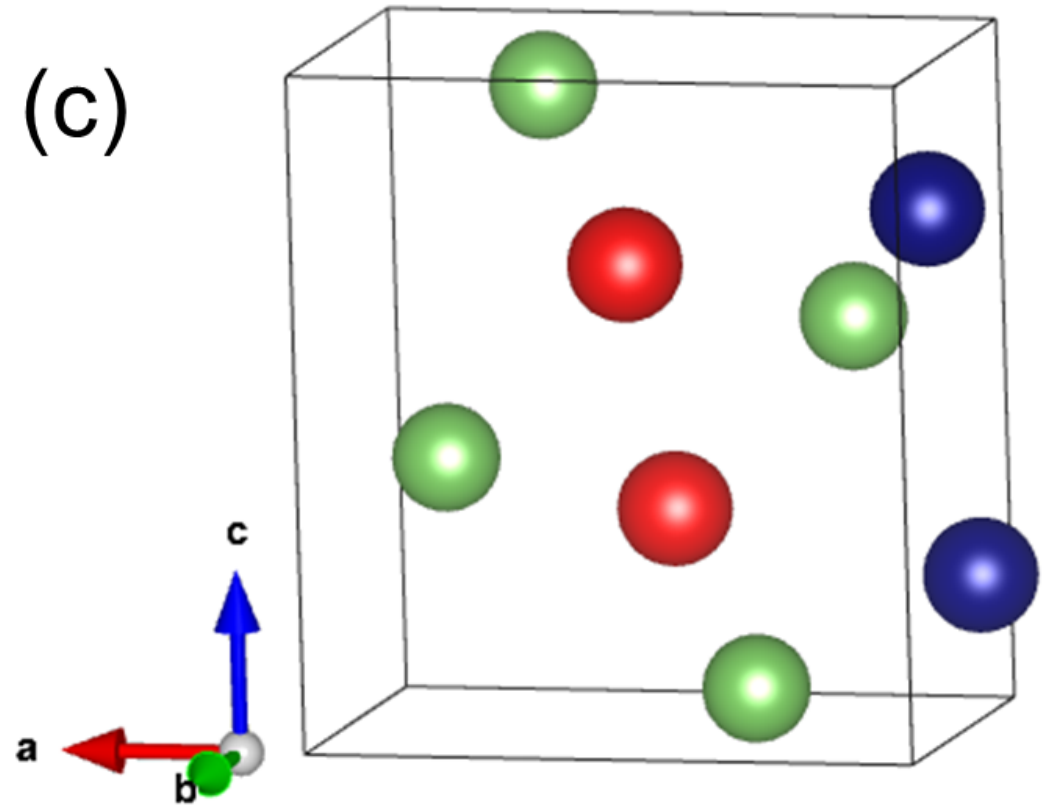}
\includegraphics[width=4.19cm,angle=0]{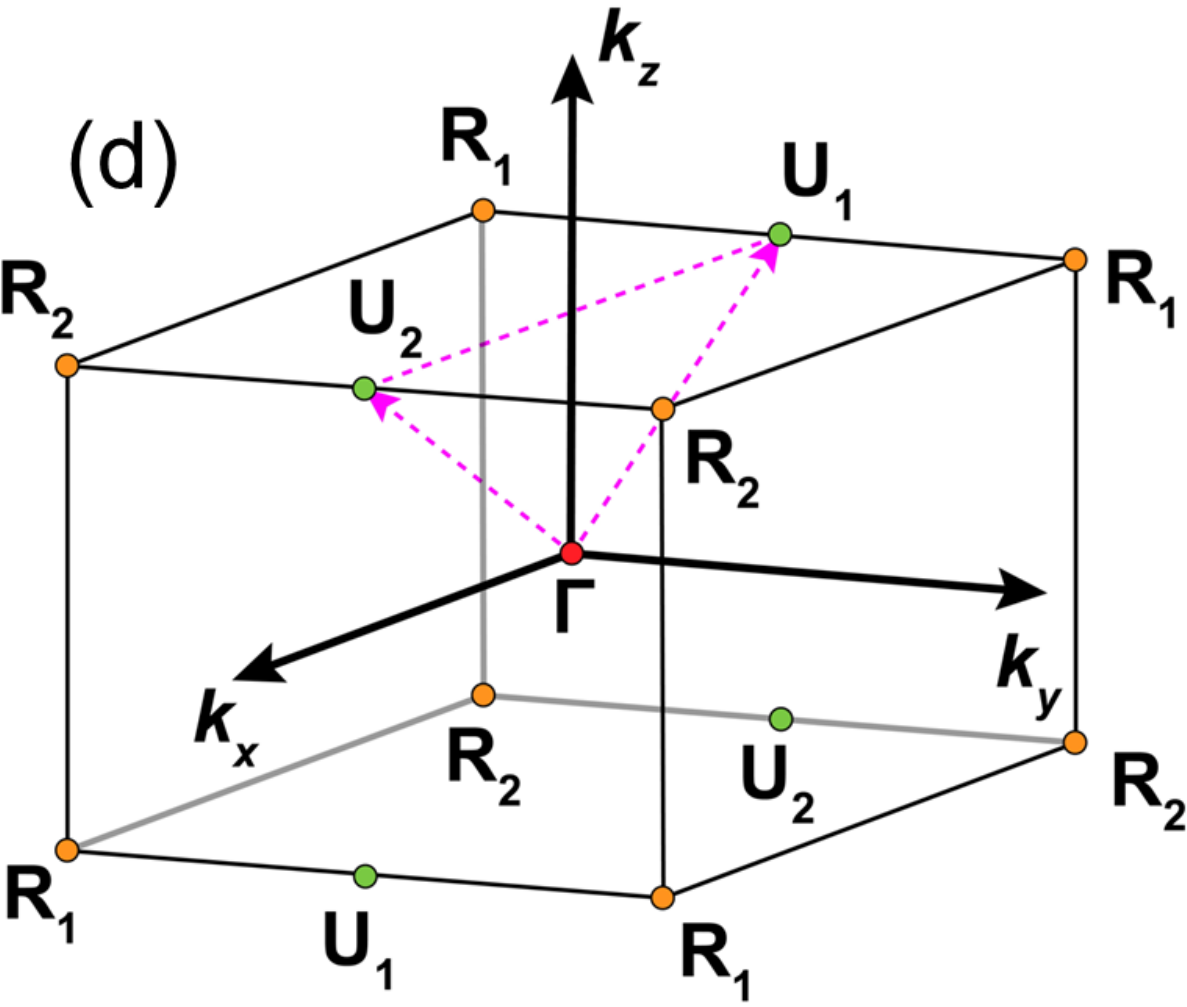}
\caption{Crystal structure and collinear magnetic orders for CrAs as (a) A-type, (b) G-type and (c) C-type. The balls with colors blue and red represent the Cr atoms with the opposite spin moments. Green balls represent the As atoms. (d) Symmetries of the irreducible Brillouin zone of the orthorhombic primitive cell for the C-type magnetic order. The C-type is the only magnetic order that can host AM. The position of the high-symmetry $k$-points U$_1$, U$_2$, R$_1$ and R$_2$ are highlighted in green and yellow. The dashed magenta line indicates the high-symmetry path U$_1$-$\Gamma$-U$_2$ which is one of the possible paths to show the AM in this magnetic space group.}
\label{magnetic_order}
\end{figure}

In this paper, we study the interplay between the altermagnetic properties of the Pnma phase and the nonsymmorphic symmetries using first-principle calculations. The computational details are reported in Appendix A. 
Investigating different collinear magnetic configurations as the A-type, G-type and C-type shown in Fig. \ref{magnetic_order}(a-c), respectively, we obtained that the altermagnetic spin-splitting is present only in the C-type configuration while it is absent in the A-type and G-type ones. Even if the C-type is not the ground state\cite{Autieri17CrAsPM}, we are interested in the interplay between altermagnetism and nonsymmorphic symmetries that is a general aspect that could appear in several other compounds.
We find that the altermagnetic spin-splitting can be up to 0.5 eV in the $d$-bands close to the Fermi level while it gets reduced to 0.2 eV in the $p$-bands. 
Furthermore, we find that the interplay between AM and nonsymmorphic symmetries produces an intricate network of crossings and anticrossings in large areas of the Brillouin zone. As a result, the presence of nonsymmorphic symmetries generates a large anomalous Hall conductivity. The paper is organized as follows: in the next Section, we show the main results, while the last Section is devoted to conclusions.

%figure1

%%%%%%%%%%%%%%%%%%%%%%%%%%%%%%%%%%%%%%%%%%%%
\section{Results}

This Section is divided into two subsections with the results without SOC in the first one and the results with SOC in the second one. We report the electronic properties focusing on the interplay between AM and nonsymmorphic symmetries by considering different magnetic orders. The semi-Dirac points generate a large number of crossings and anticrossings that we describe.
When we add the SOC in the second subsection, the calculation of the anomalous Hall conductivity (AHC) confirms a relatively large value due to the several crossings and anticrossings.

%%%%%%%%%%%%%%%%%%%%%%%%%%%%%%%%%%%%%%%%%%
\subsection{Electronic properties without SOC and nonsymmorphic symmetries}

In the nonmagnetic phase, the bands are twofold degenerate at any $k$-vector of the Brillouin zone due to the presence of the inversion-time reversal symmetry. This is the Kramers degeneracy and it is protected under the action of the spin-orbit coupling interaction. The nonsymmorphic symmetries produce additional degeneracy at the border of the Brillouin zone.
The $\Gamma$ and R points are time-reversal invariant momenta (TRIM) points.
In the Pnma space group without magnetism, the $\Gamma$ point has only the Kramer degeneracy, two additional nonsymmorphic symmetries along the SR line produce an eightfold degeneracy at
R and S, while X, Y, Z, U and T have degeneracy 4 due to one nonsymmorphic symmetry. When we consider the C-type magnetism, $\Gamma$ is a TRIM point and it still presents a twofold degeneracy. In this magnetic configuration, the TRIM points R, S and T have fourfold degeneracy due to an additional nonsymmorphic symmetry while all the other high-symmetry points have degeneracy 2.
The fourfold degeneracy at the TRIM point R makes this space group one of the few where we can study the coexistence of AM and nonsymmorphic symmetries. 

\begin{figure}[t!]
\centering
\includegraphics[width=6.3cm,height=9.0cm,angle=270]{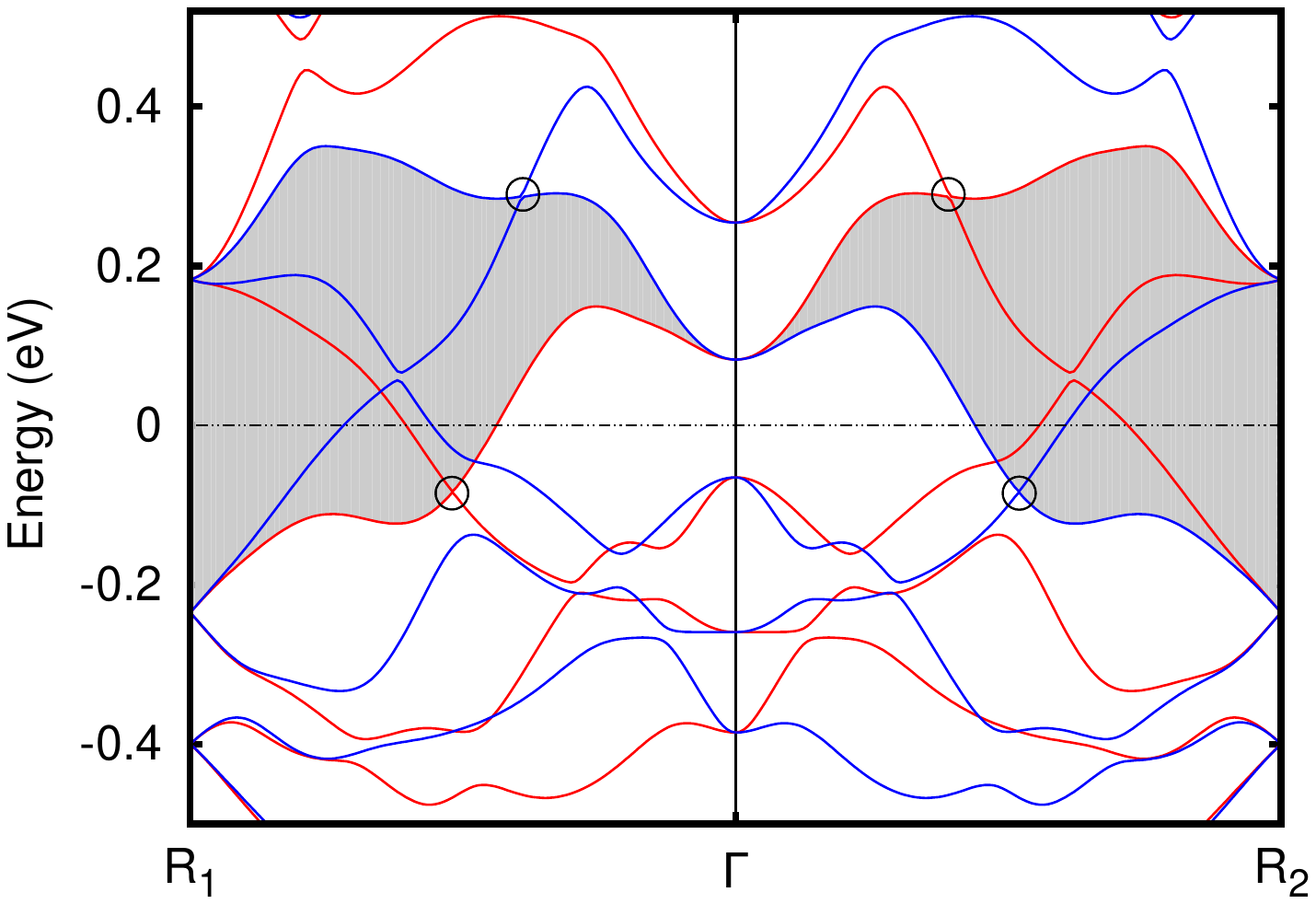}
\caption{Band structure of the C-type magnetic order along the $k$-path R$_1$-$\Gamma$-R$_2$. The spin-up channel is shown in blue, while the spin-down channel is shown in red. 
The grey area represents the non-relativistic spin-splitting. 
The band structure is plotted between -0.5 and +0.5 eV where the $d$-electrons dominate. The black circles represent the band crossings protected by glide symmetry.}\label{altermagnetism_dbands}
\end{figure}%figure2

\begin{figure}[t!]
\centering
\includegraphics[width=6.3cm,height=9.0cm,angle=270]{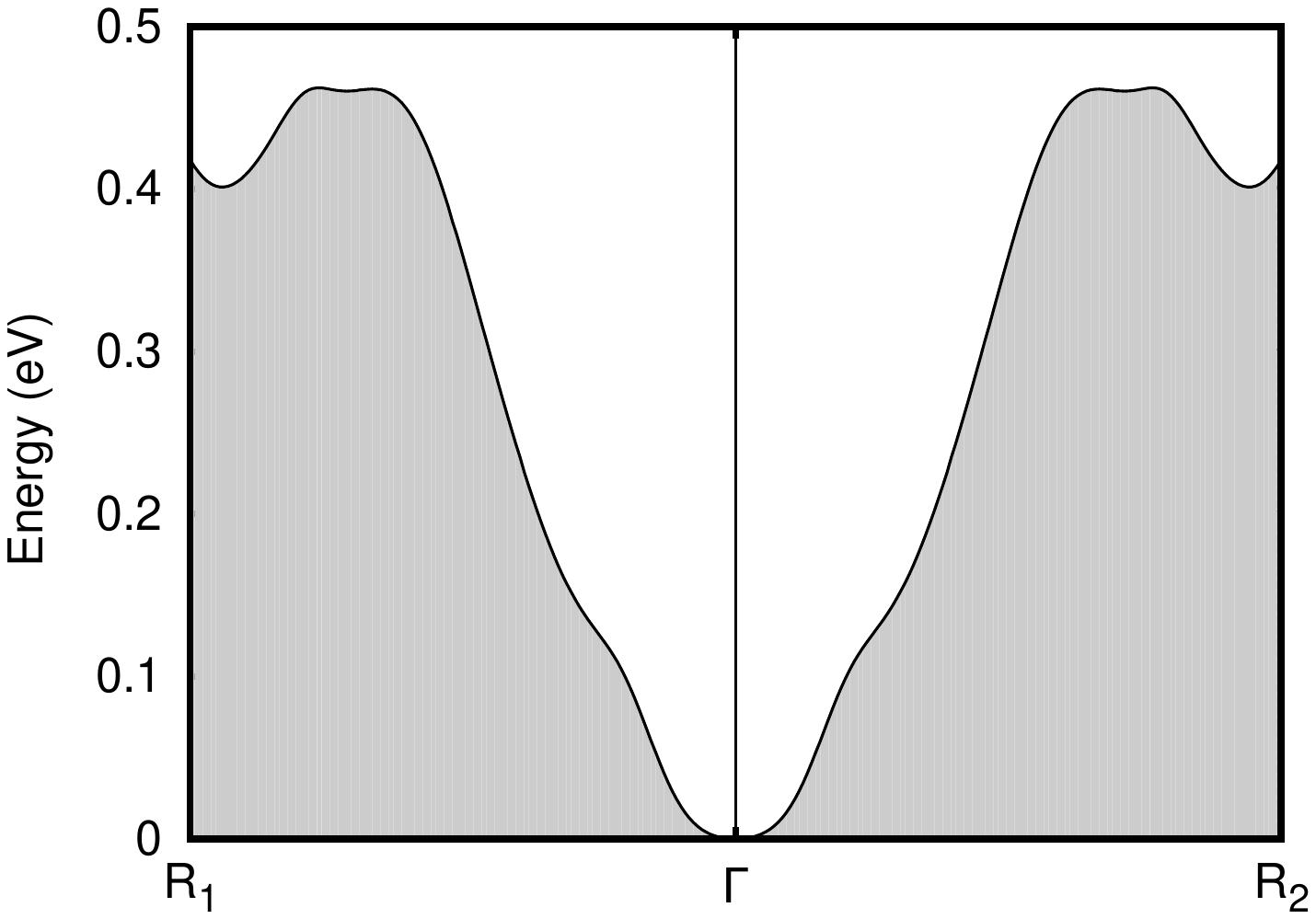}
\caption{Nonrelativistic spin-splitting along the R$_1$-$\Gamma$-R$_2$ for the first bands above the Fermi level at the $\Gamma$ point. The spin-splitting reaches the value of 0.46 eV. For every couple of bands producing a finite spin-splitting at the TRIM point, another couple of bands will produce an opposite spin-splitting.}\label{spinsplitting}
\end{figure}%figure3

\begin{figure}[t!]
\centering
\includegraphics[width=8.4cm,height=8.5cm,angle=0]{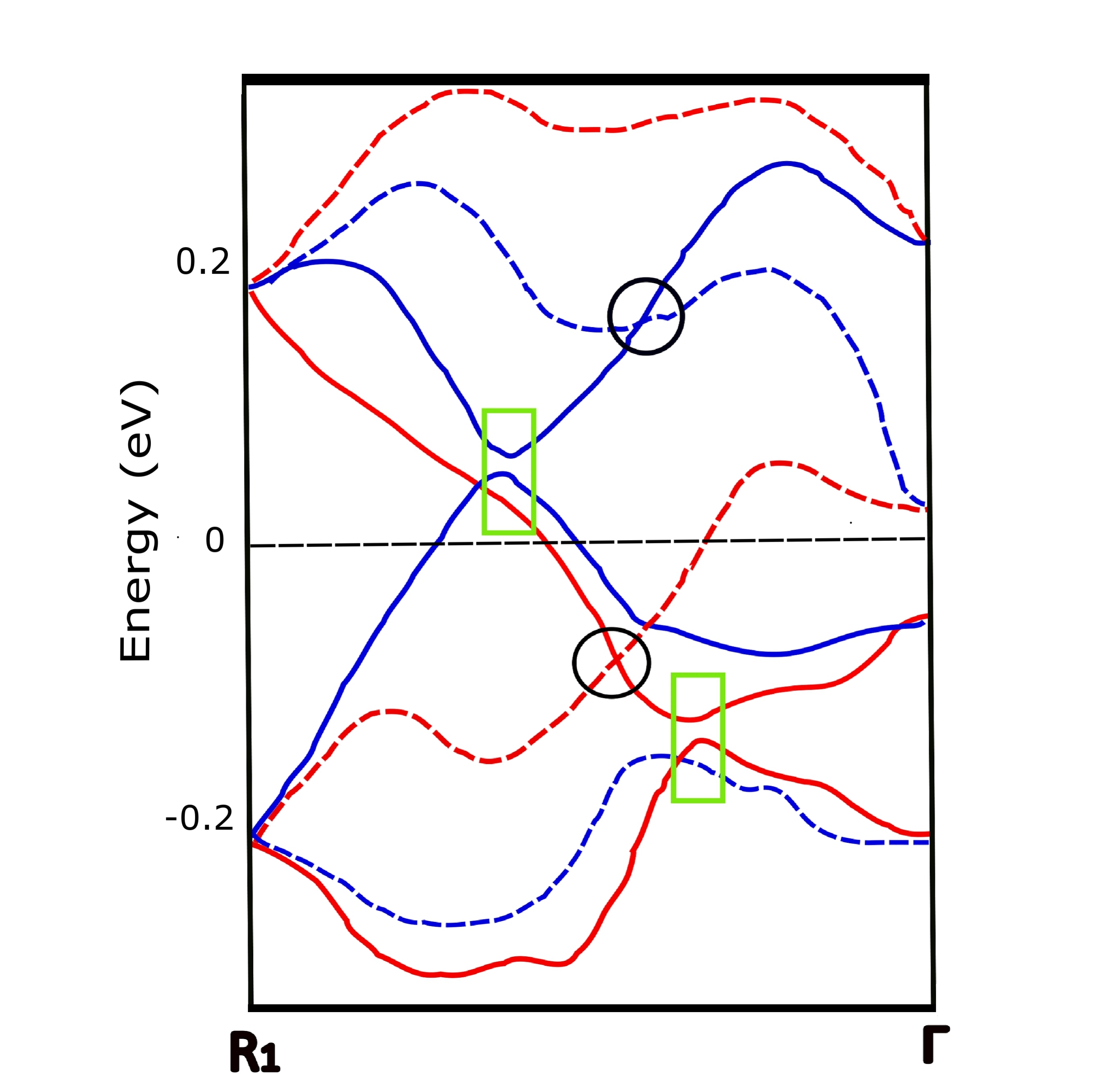}
\caption{Schematic representation of the crossings and anticrossings network in altermagnetic semi-Dirac fermions without SOC. The spin-up channel is shown in blue, while the spin-down channel is shown in red. The solid line is used for the positive eigenvalues of the glide operator, while the dashed line is used for the negative eigenvalue of the glide operator. The black circles represent the band crossings between band with opposite glide eigenvalues, while the green boxes represent the band anticrossings between bands with the same glide eigenvalues.}\label{tree_nosoc}
\end{figure}%figure4

\begin{figure*}[t!]
\centering
\includegraphics[width=5.1cm,height=5.9cm,angle=270]{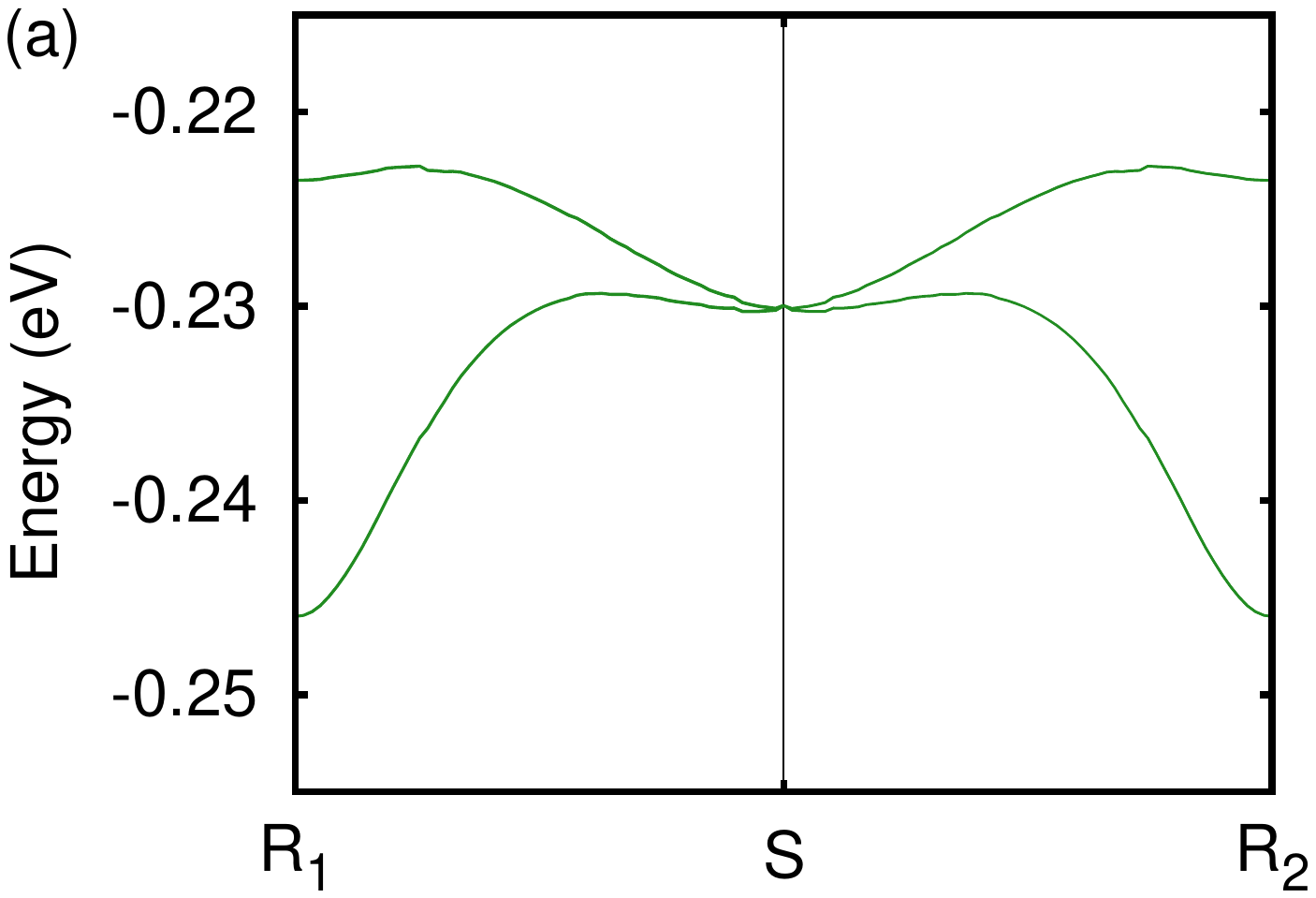}
\includegraphics[width=5.1cm,height=5.9cm,angle=270]{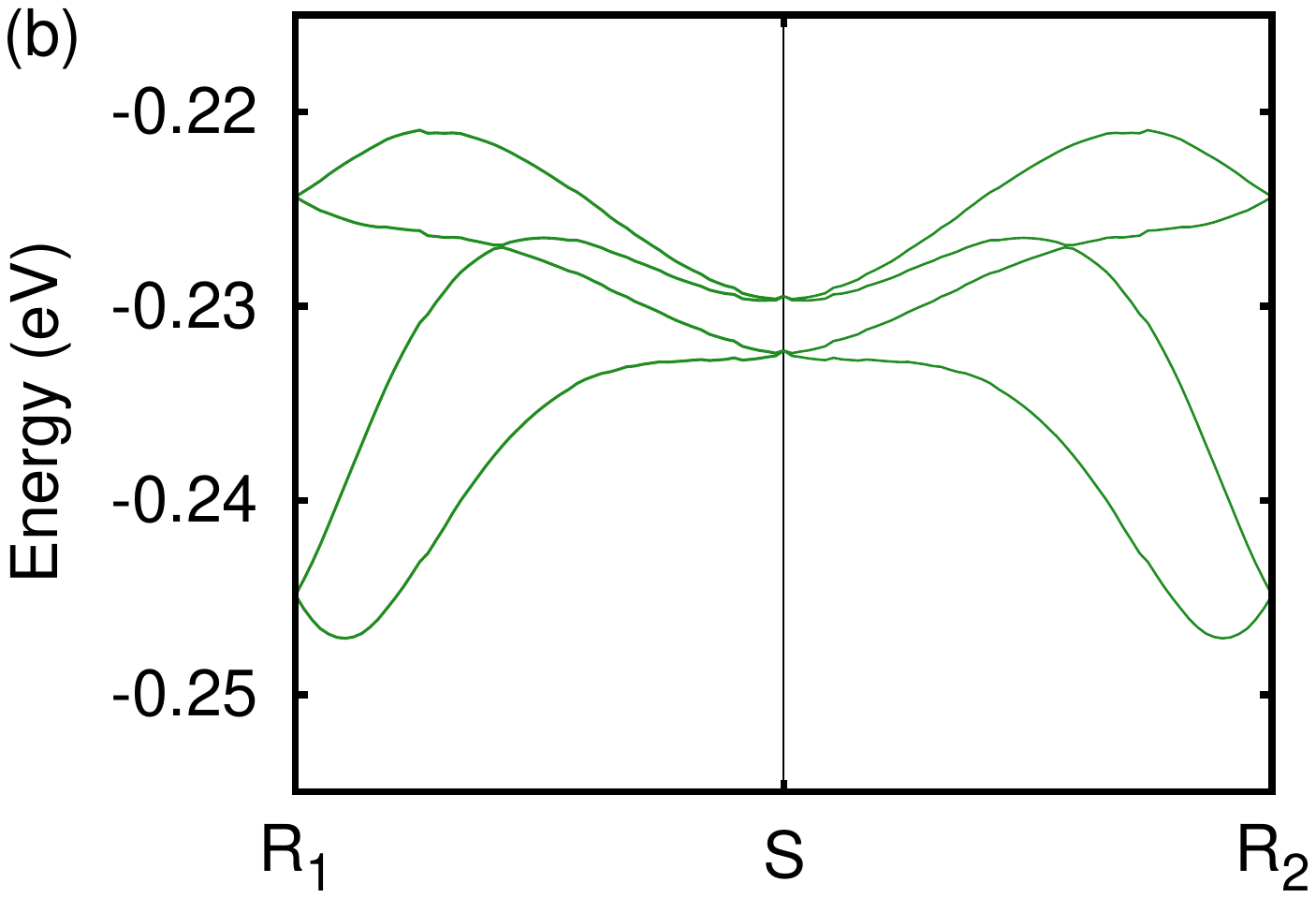}
\includegraphics[width=5.1cm,height=5.9cm,angle=270]{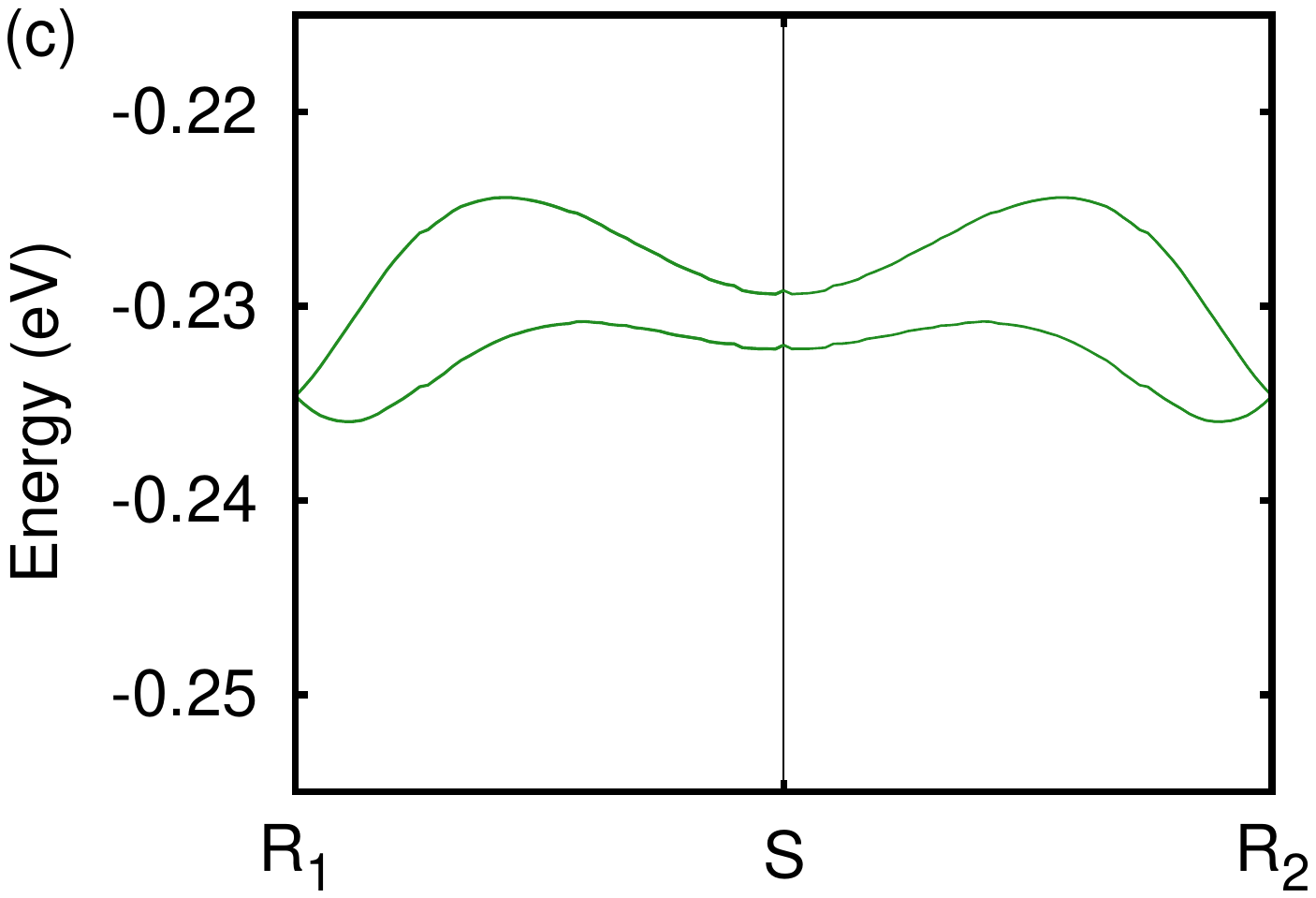}
\caption{Band structure along the R$_1$-S-R$_2$ $k$-path for the C-type magnetic order including SOC with the N\'eel vector along the (a) x-axis, (b) y-axis and (c) z-axis. No AM is present along this $k$-path. In panel (b), we obtain the antiferromagnetic hourglass fermions with a magnetic space group type-I. No hourglass fermions are found in the magnetic space groups of type-III.}\label{SOC_dbands_RS}
\end{figure*}%figure5

%Missing altermagnetism in A-type and G-type magnetic configuration
We find AM in the C-type magnetic order, but not in the other magnetic orders and this is due to the following reasons.
In the space group no. 62, there are 16 different magnetic space groups. 8 of these magnetic space groups can host AM, while the other 8 groups cannot.\cite{Gallego:db5106} In the C-type magnetic order, the space groups that depend on the N\'eel vector belong to type-I and type-III. Both of these types present AM. Changing the magnetic configuration to G-type or A-type, the magnetic space group changes to type-II or type-IV, therefore, different magnetic configurations can have different altermagnetic properties. 
%In particular, in the Pnma space group a nonsymmorphic symmetry composed of two mirrors along the x=0 plane and z=0 plane combined with a half-vector translation along the y-axis is present. This nonsymmorphic symmetry maps the up-spins in down-spins and vice versa for the A-type and G-type magnetic orders, therefore no roto-translation is needed to map spin-up in spin-down. On the contrary for the C-type, the same nonsymmorphic symmetry map a spin-up (down) site in the other spin-up (down) site. Therefore, the C-type is altermagnetic.
The CrAs contains 4c Wyckoff positions for both Cr and As, therefore the altermagnetic properties of the Pnma CrAs are different from the Pnma perovskites that have the magnetic atoms in 4b Wyckoff positions.\cite{PhysRevB.107.155126,Cuono23orbital}
A spin symmetry group analysis \cite{PhysRevX.12.021016} would provide additional understanding of the system.

%%%%%%%%%%%%%
The BZ for the C-type magnetic order is reported in Fig. \ref{magnetic_order}(d). With subscripts 1 and 2, we indicate the two points in the $k$-space that have opposite non-relativistic spin-splitting. In this case, the $k$-path connecting the $\Gamma$ point with U and R shows altermagnetic spin-splitting. In this paper, we will take as an example for the discussion the altermagnetic properties along the path R$_1$-$\Gamma$-R$_2$. Similar arguments will be valid for the altermagnetic properties along U$_1$-$\Gamma$-U$_2$, however, the presence of multiple $k$-paths where the AM is present means that there is a large area of the k-space where the altermagnetic spin-splittings reside.

%%%%% EQUATION SEMIDIRAC POINTS, RELEVANCE OF DIRAC CRUCIAL FOR THE PAPER
We stress the relevant role of the semi-Dirac points in the generation of the band crossings and anticrossings. We focus on the semi-Dirac points at the R point, but they are present at all borders of the Brillouin zone.
From the nonsymmorphic symmetries in space group no. 62, we obtain that the energy spectrum at the point R=($\pi$,$\pi$,$\pi$) is always a semi-Dirac point with the dispersion relations for spin-up and spin-down as :
\begin{equation}
    E_\uparrow(\pi-\epsilon,\pi-\epsilon,\pi-\epsilon)= \varepsilon_0 \pm v\epsilon
\end{equation}
\begin{equation}
    E_\downarrow(\pi-\epsilon,\pi-\epsilon,\pi-\epsilon)= \varepsilon_0 \pm v\epsilon
\end{equation}
with $\varepsilon_0$ being a combination of the on-site energies
and $v$ a combination of the first-neighbor hopping parameters\cite{Cuono19PRM}. Since the Dirac velocity does not depend on the energy difference between majority and minority electrons, it is the same for the spin-up or spin-down channel.  
In the case of large hybridization between the $p$-$d$ electrons, as happens in CrAs, the Dirac velocities are also large.
The presence of multiple orbitals and semi-Dirac bands with large Dirac velocities, which can be positive and negative,  favors the creation of several band crossings and band anticrossings which are observable in the band structure as presented in Fig. \ref{altermagnetism_dbands}. 
What we have shown in Fig. \ref{altermagnetism_dbands} for the Cr-$d$ bands close to the Fermi level happens qualitatively also for the As-$p$ bands as we have described in Appendix B.
The bands with opposite Dirac velocity +$v$ and -$v$ also have opposite glide eigenvalues and orthogonal eigenvectors.
Therefore, the bands with opposite glide eigenvalues are orthogonal and do not hybridize each other\cite{PhysRevB.104.125135,PhysRevB.103.155144,Zhao16}, this is exactly true for the bands with linear dispersion very close to the border of the Brillouin zone, while away from the Brillouin zone border the bands tend to show mixing of the eigenvectors. 
In Fig. \ref{altermagnetism_dbands}, we have reported with black circles the band crossings protected by the glide symmetry.

%%%%%%% SPIN SPLITTING

Now, we move to the analysis of the nonrelativistic spin-splitting in the presence of nonsymmorphic symmetries.
If we consider the Kramers pair bands which at the $\Gamma$ point are at 0.1 eV above the Fermi level and follow them, we can see that the spin-up and spin-down channels have both a glide-protected band crossing and then they reach the R points at two different eigenvalues creating surprisingly a finite nonrelativistic spin-splitting at a TRIM point as shown in Fig. \ref{spinsplitting}. 
The nonzero value of the spin-splitting at the point R apparently contradicts the concept of TRIM point, but there would be another couple of bands that would produce the opposite nonrelativistic spin-splitting to preserve the total spin-splitting at a given TRIM point.
Due to these exceptional conditions, the nonrelativistic spin-splitting for those bands, reported in Fig. \ref{spinsplitting}, reaches a maximum of 0.46 eV.

%%%%%%%%%% avoided crossing=ANTICROSSINGS AND CROSSINGS
In Fig. \ref{tree_nosoc}, we plot a schematic figure with the spin-channels in blue and red colors, with the solid (dashed) line used for the positive (negative) eigenvalues of the glide operator. We have the expected intrachannel avoided band crossings and the interchannel band crossings. Additionally, we have intrachannel band crossings, namely crossings between bands with opposite glide eigenvalues. In the next subsection, we will deliberate on how these crossings will evolve with SOC as a function of the N\'eel vector orientation.

\begin{figure*}[t!]
\centering
\includegraphics[width=5.9cm,height=5.9cm,angle=360]
{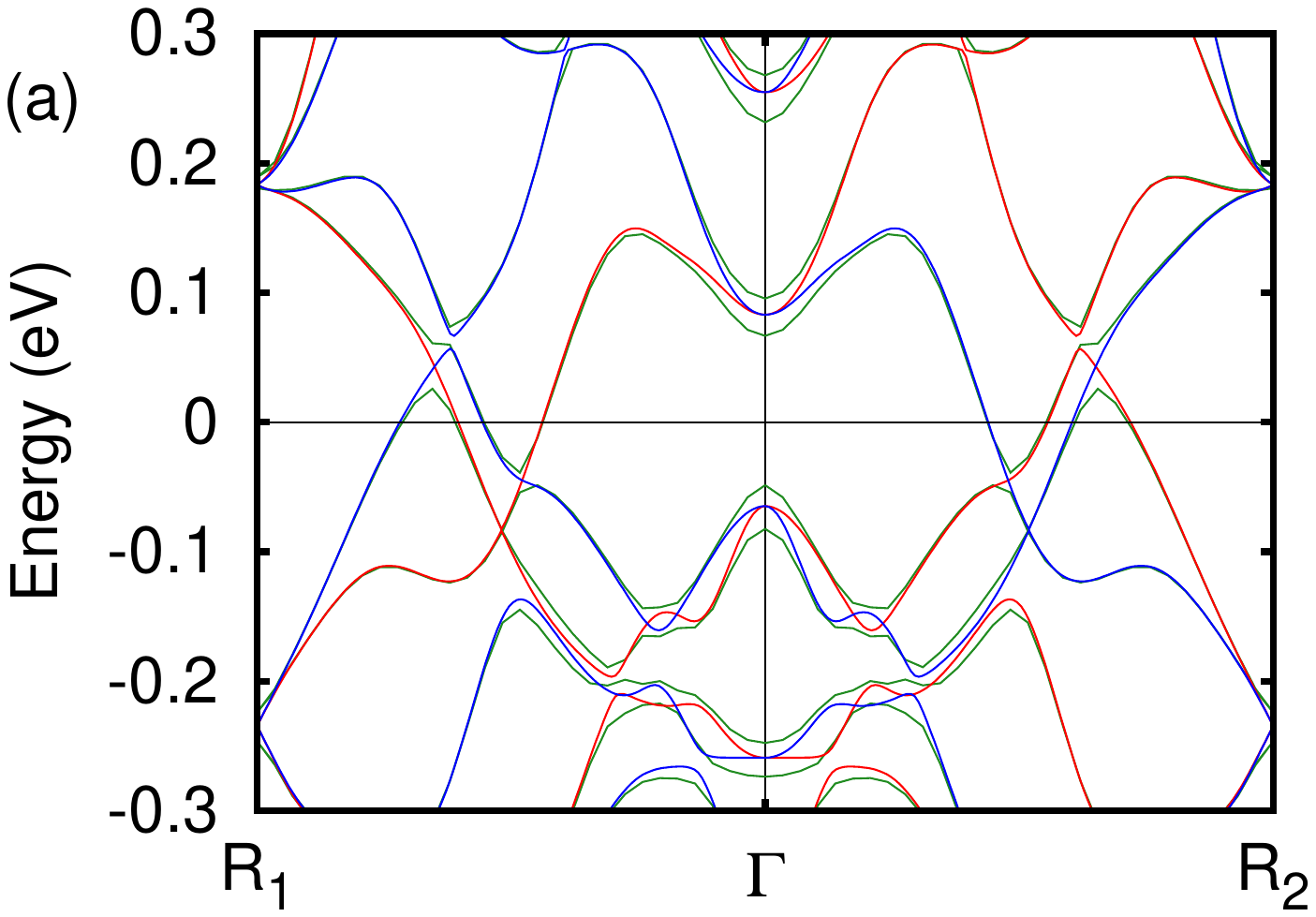}
\includegraphics[width=5.9cm,height=5.9cm,angle=360]
{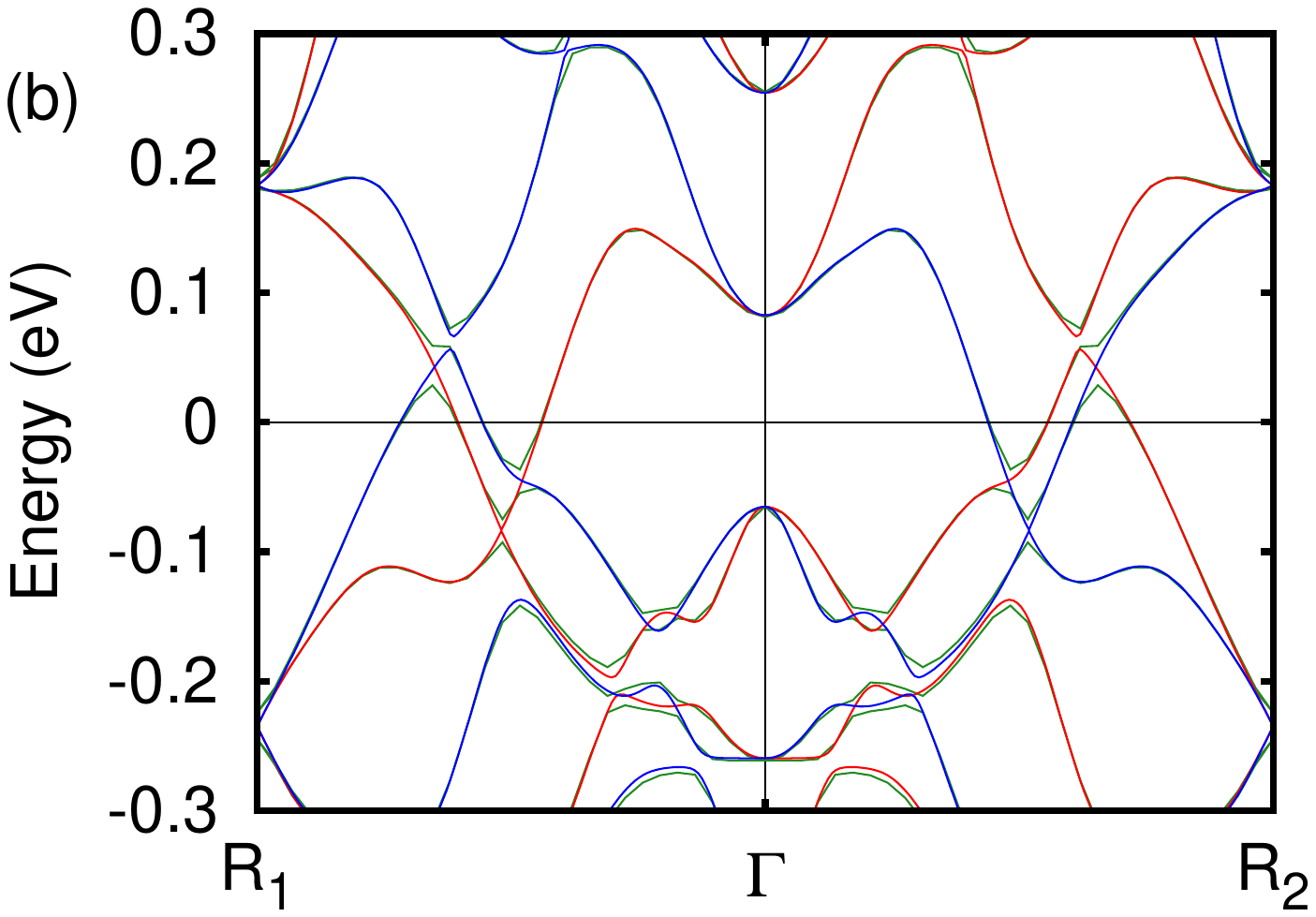}
\includegraphics[width=5.9cm,height=5.9cm,angle=360]
{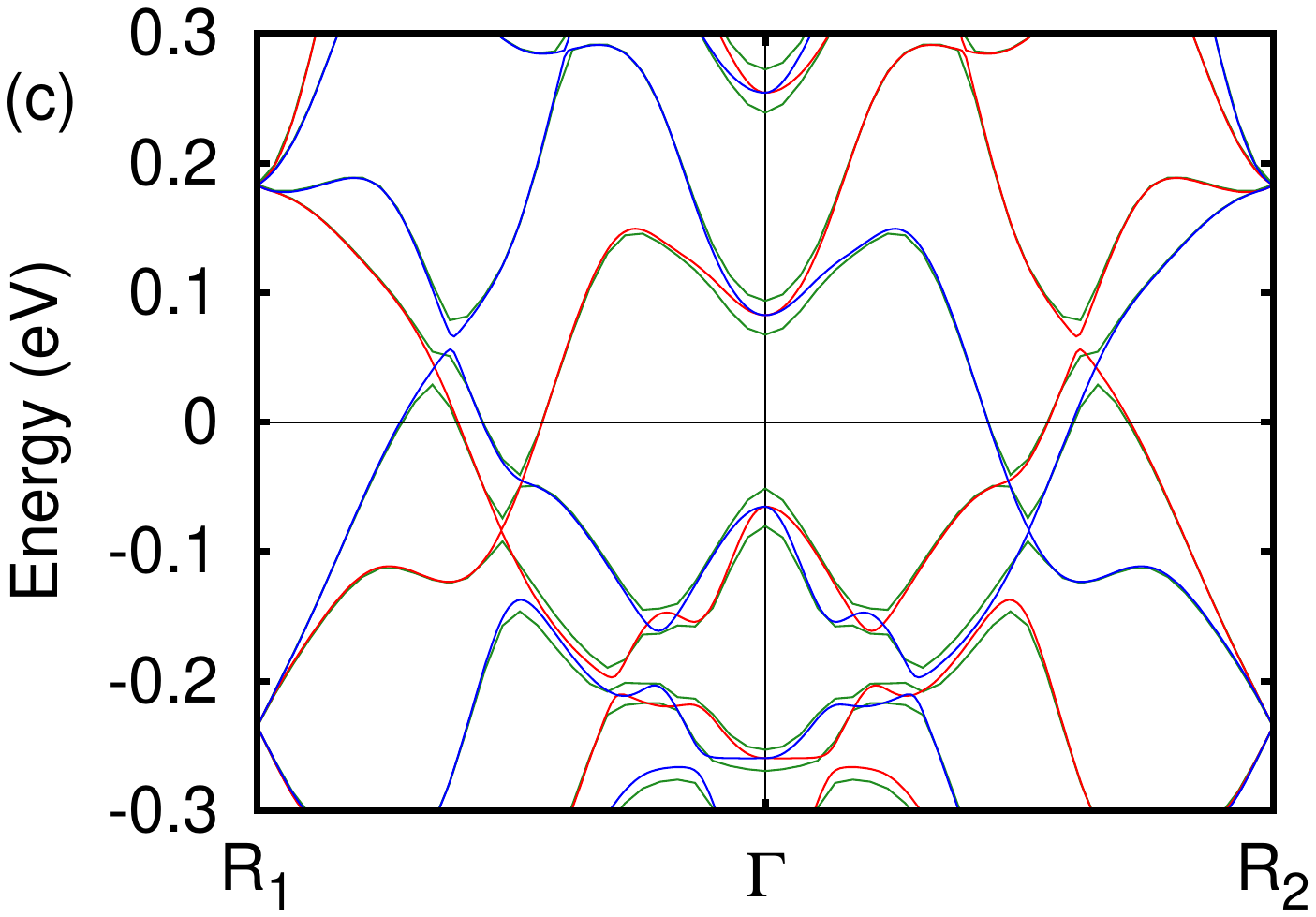}
\caption{Band structure of the C-type magnetic order along the $k$-path R$_1$-$\Gamma$-R$_2$ with N\'eel vector along the (a) x-axis, (b) y-axis and (c) z-axis, respectively. The spin-up channel is shown in blue, the spin-down channel is shown in red while the band structure with SOC is plotted in green. The band structure is plotted between -0.3 and +0.3 eV where the $d$-electrons dominate.}\label{SOC_dbands}
\end{figure*}%figure6

\begin{figure*}[t!]
\centering
\includegraphics[width=5.9cm,height=5.9cm,angle=360]
{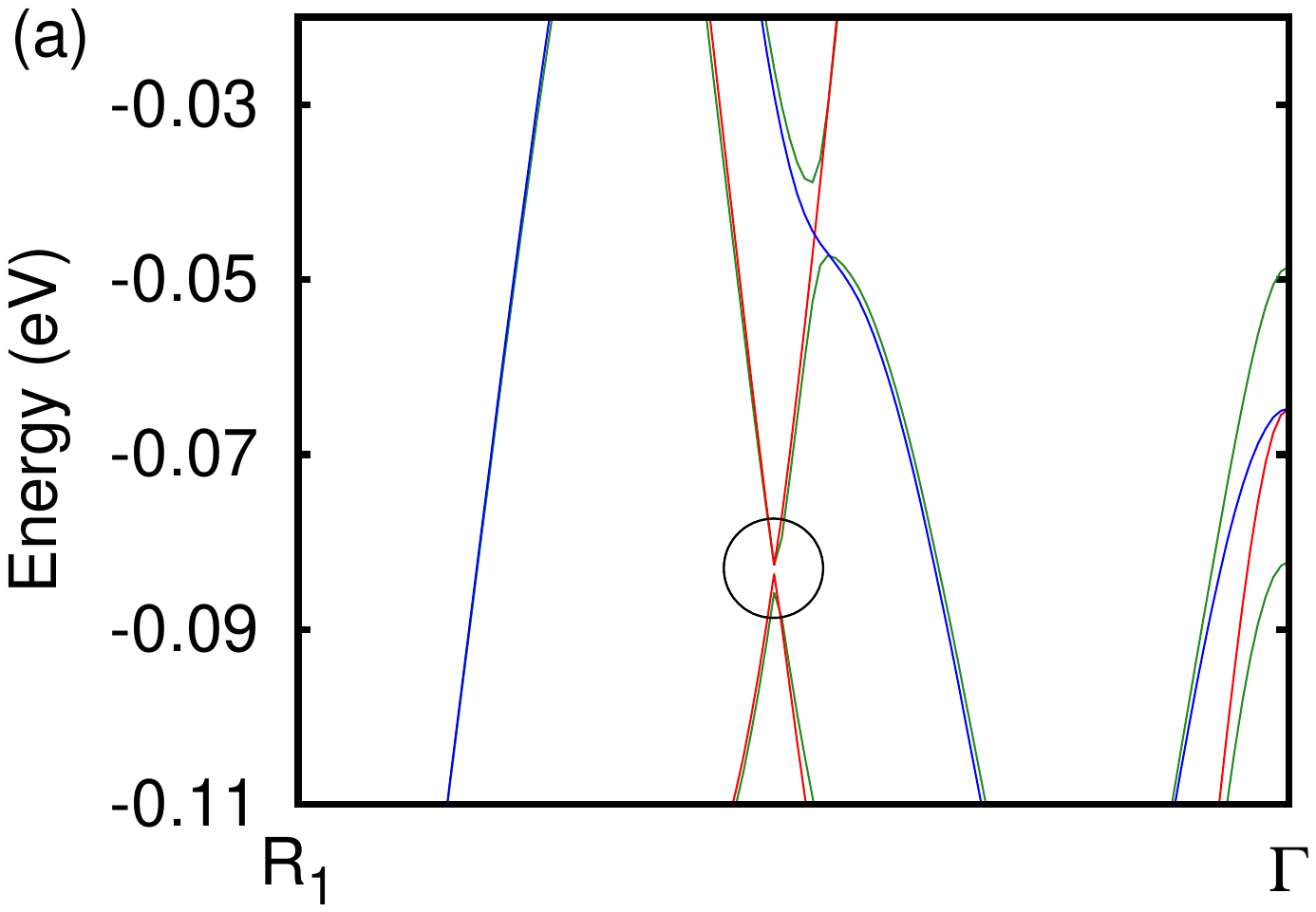}
\includegraphics[width=5.9cm,height=5.9cm,angle=360]
{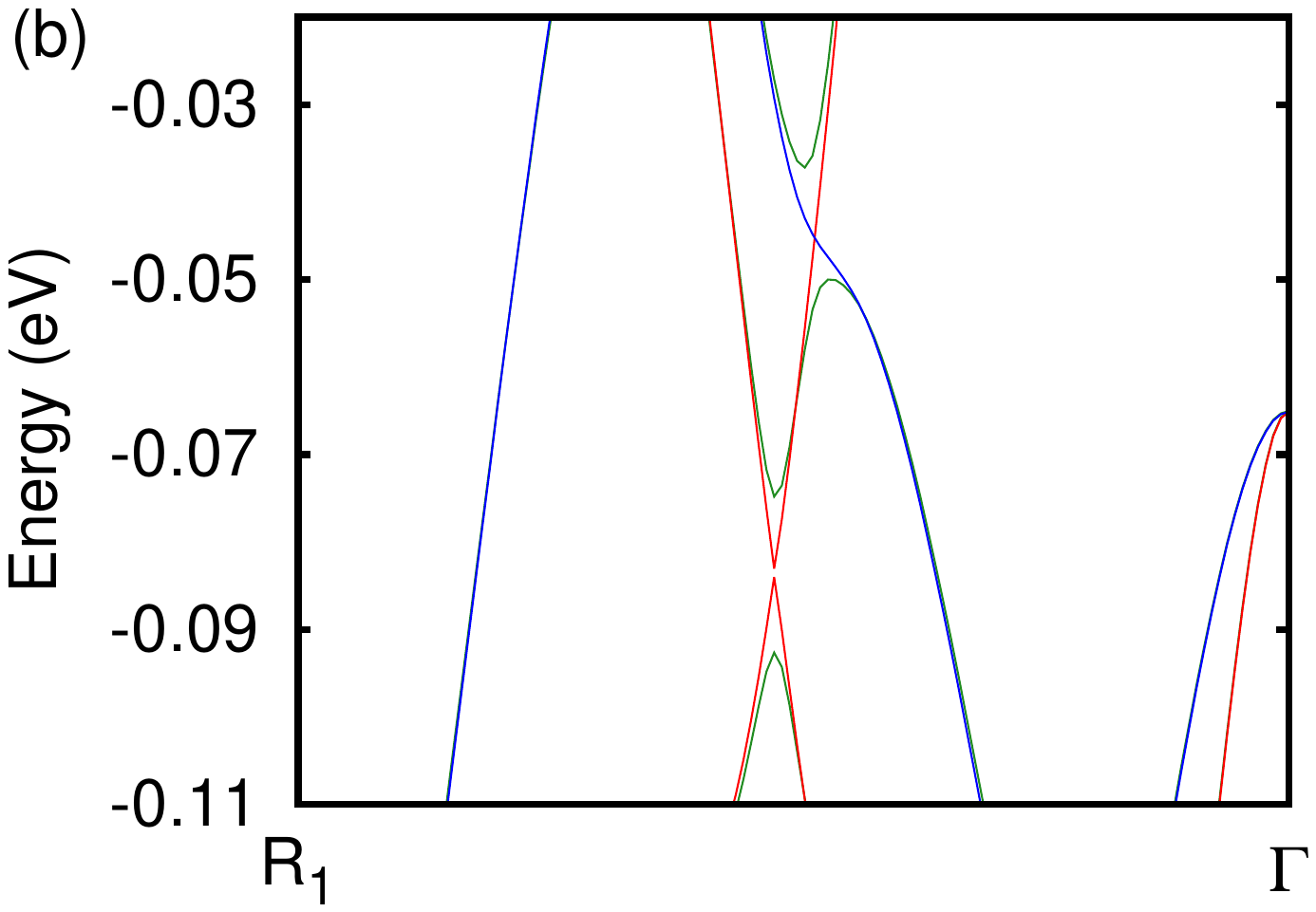}
\includegraphics[width=5.9cm,height=5.9cm,angle=360]
{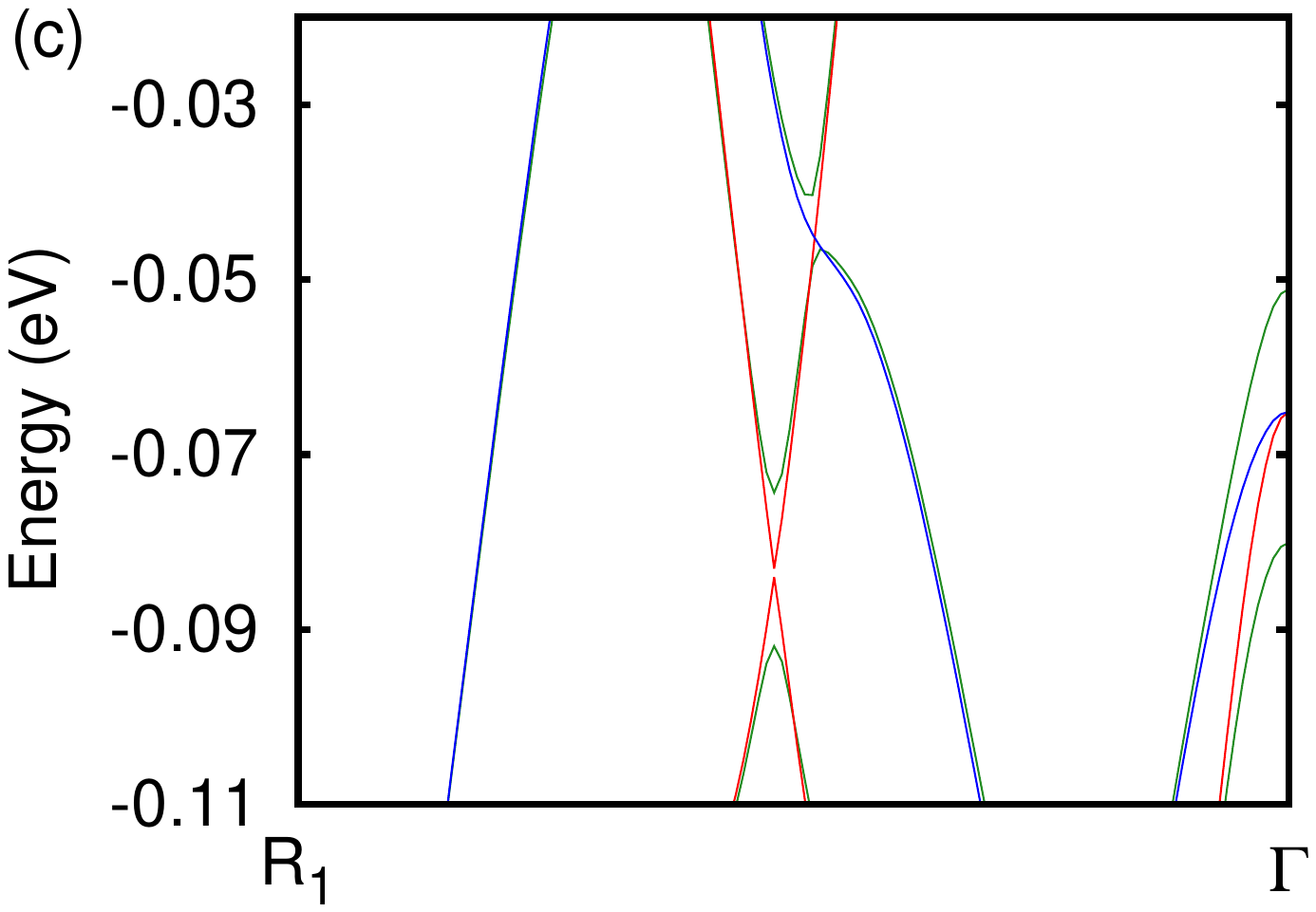}
\caption{Magnification of the band structure in FIG. 6. Band structure of the C-type magnetic order along the $k$-path R$_1$-$\Gamma$ with the N\'eel vector along the (a) x-axis, (b) y-axis and (c) z-axis. The band structure is plotted between -0.11 and -0.02 eV where there are the bands linear in $k$ with opposite glide-eigenvalues. The black circle in panel (a) describes the band crossing between opposite glide eigenvalues not splitted by SOC when the N\'eel vector is along x. }\label{SOC_dbands_mag}
\end{figure*}%figure7

%%%%%%%%%%%%%%%%%%%%%%%%%%%%%%%%%%%%%%%%%%%%%%%%%%
\subsection{Electronic properties with SOC and Anomalous Hall conductivity}

% magnetic space group

In the non-magnetic case, when we consider the SOC, we have shown, in a previous work \cite{Cuono19PRM},  a selective removal of the bands degeneracy due to the nonsymmorphic symmetries at the TRIM points and at the borders of the Brillouin zone.
In the C-type magnetic configuration, when we include the SOC interaction, we observe a selective removal of the spin-degeneracy depending on the N\'eel vector direction and consequently on the magnetic space group. Indeed, the variation of the N\'eel vector direction changes the magnetic space group. 
We define the x-, y- and z-axis as parallel to the $a$, $b$ and $c$ lattice constants, respectively, as defined in Fig. \ref{magnetic_order}.
When the N\'eel vector is along y, the magnetic space group is 62.441, which is a type I. When the magnetization is along x and z, the magnetic space groups are 62.446 and 62.447, respectively, which are type III. This selective removal of the degeneracy is valid along $k$-paths with and without AM.  
Looking at Figs. \ref{SOC_dbands_RS} and \ref{SOC_dbands}, we can see that the spin-orbit splitting is qualitatively and quantitatively different depending on the magnetic space group.

%%%% splitting due to SOC for R-S-R
It was shown that, in the non-magnetic case, the SOC acts selectively at the TRIM points and Brillouin zone border due to the nonsymmorphic symmetries\cite{Cuono19EPJST}.
We will investigate the SOC effects in the magnetic phase as a function of the N\'eel vector starting from the path  R$_1$-S-R$_2$ where no AM is present. Without SOC, the entire RS line including the TRIM points have degeneracy 4.
The band structure with SOC along the R$_1$-S-R$_2$ $k$-path is reported in Fig. \ref{SOC_dbands_RS}. When the N\'eel vector is along x, the spin-orbit splits the bands at R but not at S. When the N\'eel vector is along z, the spin-orbit splits the bands at S but not at R.
When the N\'eel vector is along y, the spin-orbit splits
the bands at both R and S, and therefore, in this latter case, we find antiferromagnetic hourglass fermions. 
These hourglass fermions are relativistic features already present along the RS line without magnetism\cite{Wang16Nature,Wang17PRB,Li18PRB}.
Recently, the ferromagnetic and antiferromagnetic hourglass fermions were investigated. However, we find the antiferromagnetic hourglass in the magnetic space group 62.441 that was not reported before\cite{PhysRevB.106.165128}.

%%%% splitting due to SOC for R-Gamma-R
The spin-orbit acts selectively also along the R$_1$-$\Gamma$-R$_2$ path where the nonrelativistic spin-splitting is present. The effects of the SOC for different N\'eel vector orientations are reported in Fig. \ref{SOC_dbands}, we will highlight in the discussion the $k$-points where the SOC is not effective. 
When the N\'eel vector is along x, we have additional SOC splittings for all the crossing and anticrossing points except that for the intrachannel crossing  point at around -0.1 eV, which is protected by the glide operator as shown in Fig. \ref{SOC_dbands}(a). A magnification of Fig. \ref{SOC_dbands}(a-c) is reported in Fig. \ref{SOC_dbands_mag}(a-c), respectively, to highlight the band crossings protected by the glide against SOC when the N\'eel vector is along x. Indeed, the band crossing in the black circle in Fig. \ref{SOC_dbands_mag}(a) is not splitted by SOC, while the SOC splitting is evident in \ref{SOC_dbands_mag}(b,c). It was shown that this kind of glide-protected band crossings can generate a non-isoenergetic nodal-line\cite{PhysRevB.103.155144,PhysRevB.104.125135}. Therefore, this case would be a nodal-line in the altermagnetic phase, further investigations in this direction using model hamiltonian calculations could be interesting for the topological matter community.  
When the N\'eel vector is along y, we do not have SOC splitting at the $\Gamma$ point as obtained in Fig. \ref{SOC_dbands}(b). When the N\'eel vector is along z as shown in Fig. \ref{SOC_dbands}(c), we do not have SOC splitting at the R$_1$ and R$_2$ points coherently with what is observed in Fig. \ref{SOC_dbands_RS}(c).

\begin{figure}[t!]
\centering
\includegraphics[width=1\linewidth]{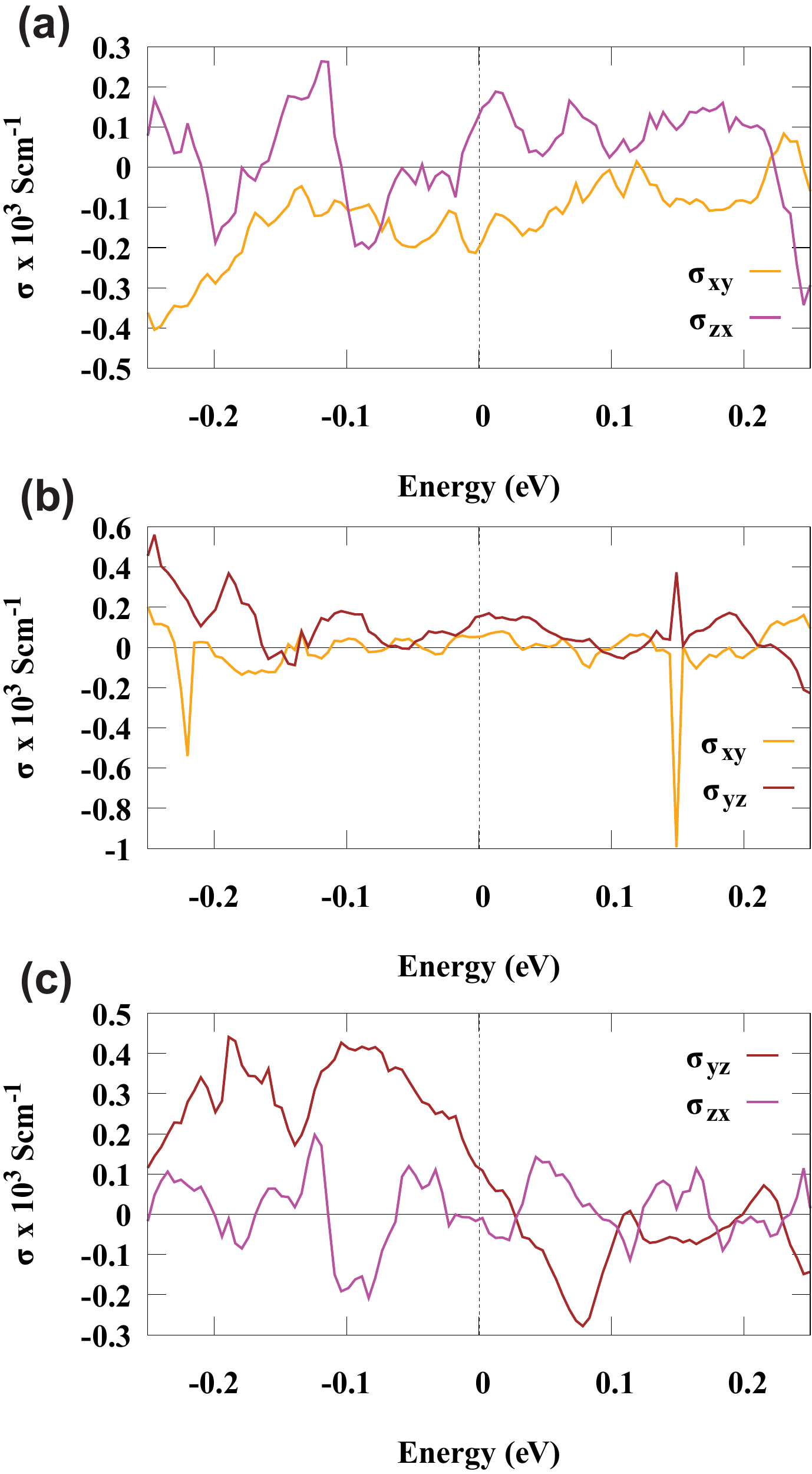}
\caption{Anomalous Hall effect in altermagnetic C-type magnetic order with N\'eel vector along x, along y and along z reported in panels (a), (b) and (c), respectively. In every case, we show the 
AHCs $\sigma_{xy}$, $\sigma_{xz}$ and $\sigma_{yz}$ for an Hall vector orthogonal to the N\'eel vector. The AHCs are presented between -0.25 and +0.25 eV, in this energy range the maximum in absolute value was reached by $\sigma_{xy}$ around -1000 Scm$^{-1}$ with N\'eel vector along y-axis. The Fermi energy is set to zero.}\label{AHE_Neel}
\end{figure}%figure8

%%%% Hall vector and AHC
The Hall vector in altermagnetic systems with space group no. 62 can lie in all possible directions in principle\cite{doi:10.1126/sciadv.aaz8809}. However, when the structural details are added, the Hall vector has a precise orientation. It was shown that in space group no. 62, we can expect the Hall vector in directions orthogonal to the N\'eel vector\cite{PhysRevB.107.155126}.
Several anticrossing points appear in the whole Brillouin zone boundary due to the presence of the semi-Dirac points. These anticrossing points generate large Berry curvature and consequently a large anomalous Hall effect, since, as it is well known from the literature \cite{Nagaosa10,Xiao10,Taguchi01,Chakraborty22,Zhang23,Jungwirth02,Yao04,Chen14,Vanthiel21}, the intrinsic anomalous Hall effect can be expressed in terms of the Berry curvature.
The semi-Dirac points and the glides linked to the nonsymmorphic symmetries are key ingredients to the generation of these several crossings and anticrossings. When we add the SOC, avoided band crossings are obtained and a large AHC is expected. Therefore, we can claim that the interplay between AM and nonsymmorphic symmetries generates large anomalous Hall conductivity.  
We calculate the AHC between -0.25 and +0.25 eV and we report it in Fig. \ref{AHE_Neel} for different N\'eel vector orientations along the principal axes. The three AHC components that we have calculated are $\sigma_{yz}$, $\sigma_{xz}$ and $\sigma_{xy}$, the numerical details are reported in Appendix A.
We obtain large values of the AHE in the two directions orthogonal to the N\'eel vector. When the N\'eel vector is along x, we obtain AHCs values up to -400 S/cm for the $\sigma_{xy}$ and 300 S/cm for the $\sigma_{xz}$ (see Fig. \ref{AHE_Neel} (a)).
When the N\'eel vector is along y, we obtain AHC values up to -1000 for the $\sigma_{yz}$ and +600 S/cm for the $\sigma_{xy}$ (see Fig. \ref{AHE_Neel} (b)), the spike that produces the large values at +0.14 eV was verified with a denser energy grid.
Finally, when the N\'eel vector is along z, we obtain AHC values up to 450 for the $\sigma_{yz}$ and -200 S/cm for the $\sigma_{xz}$ (see Fig. \ref{AHE_Neel} (c)).
We observe a strong change of all the AHCs when we switch the N\'eel vector from the one axis to another.
The compounds with space group no. 62 and magnetic atoms in Wyckoff position 4b hosts the AHC in one component\cite{PhysRevB.107.155126}, however, compounds with the same space group but different Wyckoff positions host an Hall vector orthogonal to the N\'eel vector. 
These values are smaller but of the same order of magnitude as the AHC in other altermagnetic metallic compounds\cite{doi:10.1126/sciadv.aaz8809,PhysRevB.107.155126} as RuO$_2$ and CaCrO$_3$.

%%%%%%%%%%%%%%%%%%%%%%%%%%%%%%%%%%%%%%%
\section{Conclusions}

The presence of AM varies with the magnetic configuration, since the magnetic space group type strongly depends on the magnetic configuration. For the space group no. 62 with magnetic atoms in 4c Wyckoff position, the non-relativistic spin-splitting is present for the C-type magnetic order while it is absent in G-type and A-type magnetic orders. Following a couple of bands with opposite spin from the degenerate $\Gamma$ point through the glide-protected crossing, we can end up with a finite spin-splitting at the TRIM points R$_1$ and R$_2$, however, there would be another couple of bands that would produce the opposite spin-splitting restoring the zero non-relativistic spin-splitting at the TRIM points. 
The magnetic space group is type-I when the N\'eel vector is along y and type-III when the N\'eel vector is along x or z. A selective removal of the spin degeneracy acts as a function of the magnetic space group at the TRIM points. For example, in the type-I magnetic space group, we find antiferromagnetic hourglass electrons that are not present when the N\'eel vector is along x or z. When the N\'eel is along x, we have a glide-protected crossing that could generate a nodal-line in the altermagnetic phase.
Due to the semi-Dirac points and glide symmetries, the interplay between AM and nonsymmorphic symmetries produces several band crossings and avoided band crossings, once we apply the SOC these crossings generate a large AHC of the same order of magnitude found in altermagnetic RuO$_2$ and CaCrO$_3$.\\

%%%%%%%%%%%%%%%%%%%%%%%%%%%%%%%%%%%%%%%%%%%%%
\begin{acknowledgments}
We thank Tomasz Dietl, Canio Noce and Rajibul Islam for useful discussions.
The work is supported by the Foundation for Polish Science through the International Research Agendas program co-financed by the European Union within the Smart Growth Operational Programme (Grant No. MAB/2017/1). A.F. was supported by the Polish National Science Centre under Project No. 2020/37/B/ST5/02299.
We acknowledge the access to the computing facilities of the Interdisciplinary Center of Modeling at the University of Warsaw, Grant g91-1418, g91-1419 and g91-1426 for the availability of high-performance computing resources and support.
We acknowledge the CINECA award under the ISCRA initiative  IsC99 "SILENTS”, IsC105 "SILENTSG" and IsB26 "SHINY" grants for the availability of high-performance computing resources and support. We acknowledge the access to the computing facilities of the Poznan Supercomputing and Networking Center Grant No. 609.\\
\end{acknowledgments}

\begin{figure}[t!]
\centering
\includegraphics[width=6.8cm,height=9.1cm,angle=270]{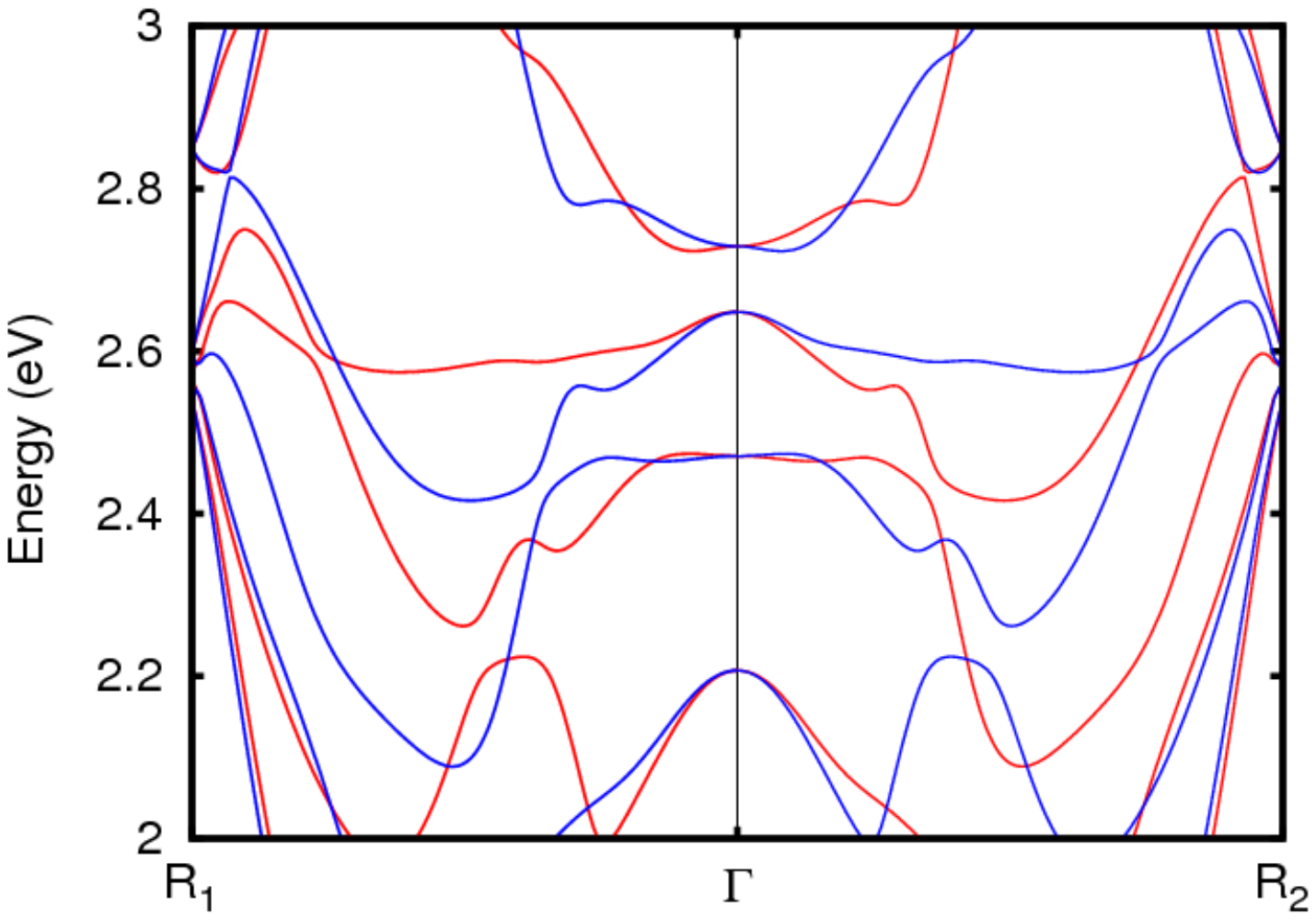}
\caption{Band structure of the C-type magnetic order along the $k$-path R$_1$-$\Gamma$-R$_2$. The spin-up channel is shown in blue, while the spin-down channel is shown in red. The band structure is plotted between +2 and +3 eV where the As $p$-electrons dominate.}\label{altermagnetism_pbands}
\end{figure}%Figure9Appendix

\appendix

\section{Computational details}

We performed density functional theory (DFT) calculations by using the VASP package\cite{Kresse93,Kresse96,Kresse96b}. 
We employed the local density approximation and the Perdew–Zunger\cite{Perdew81} parametrization of the Ceperly-Alder data\cite{Ceperley80}.
The same parameters of previous works\cite{Autieri17CrAsPM} as the lattice constants and atomic positions of Ref. \onlinecite{Shen16} have been used. The lattice constants are a=5.60499 {\AA}, b=3.58827 {\AA} and c= 6.13519 {\AA}, while the Wyckoff positions both in 4c are Cr: (0.0070, $1/4$,
0.2049) and As: (0.2045, $1/4$, 0.5836).
The band structure plots were obtained with 130 $k$-points for every path. 
Close to the Fermi level we have the Cr $d$-orbitals sandwiched between As $p$-orbitals\cite{Autieri17CrAsPM}.
We performed the wannierization\cite{Marzari97,Souza01} by using the WANNIER90 code\cite{Mostofi08} and we considered the Cr-3$d$ and As-4$p$ orbitals as we have already done in previous papers\cite{Autieri17JPCM,Autieri18JPCS,Cuono19EPJST,Cuono19PRM,Cuono18AIP,Cuono19NPJ}. 
For the calculation of the Anomalous Hall conductivity we used WannierTools code,\cite{WU2017} we have used a $k$-grid 200 $\times$ 200 $\times$ 200. The AHC calculations were verified with a $k$-grid of 300 $\times$ 300 $\times$ 300 that reproduces the same results with negligible differences\cite{Alam23}. The information about the magnetic space groups were extracted by the 
Bilbao Crystallographic Server\cite{Gallego:db5106}.\\

\section{Induced altermagnetism in p-bands}

As mentioned in the main text, the system hosts a strong $p$-$d$ hybridization\cite{Cuono19PRM}, therefore, the AM can be induced from the Cr $d$-bands to the As $p$-bands. The $d$-bands dominate up to 1.5 eV above the Fermi level\cite{Cuono19EPJST}, while above 2 eV the bands are mainly composed by As-4$p$ spectral weight.
The altermagnetic spin-splitting is of the order of 0.5 eV on the magnetic $d$-bands close to the Fermi level and survives in the $p$-bands through the $d$-$p$ hybridization but it gets reduced. The largest spin-splitting in the $p$-orbitals is around 0.2 eV as we can see in Fig. \ref{altermagnetism_pbands}.
The Cr $d$-bands induce AM in As $p$-bands, this is slightly different from the AM in EuIn$_2$As$_2$ where the magnetic $f$-bands of Eu induce AM on the $d$-bands of the same Eu atoms\cite{cuono2023abinitio}.

\bibliography{altermagnetism}
\end{document}